\documentclass[a4paper, amsfonts, amssymb, amsmath, reprint, showkeys, nofootinbib, twoside]{revtex4-1}
\usepackage[english]{babel}
\usepackage[colorinlistoftodos, color=green!40, prependcaption]{todonotes}
\usepackage{lipsum}  
\usepackage{amsmath}  
\usepackage{amsfonts}
\usepackage{float}

\makeatletter
\def\Dated@name{Date : }%

\makeatother
\begin{document}

\title{3D tracking of Plankton with single-camera stereoscopy}
\author{J. Moscatelli}
\author{X. Benoit Gonin}
\author{F. Elias}

\affiliation{PMMH, CNRS, ESPCI Paris, Université PSL, Sorbonne Université, Université Paris Cité, F-75005, Paris, France
}
\date{\today}

\begin{abstract}

We introduce a device developed to perform a 3D tracking of passive or active particles under flow, confined in a medium of hundreds micrometers wide. Micro-objects are placed inside a vertical glass capillary and two mirrors are set behind it with a certain angle, making it possible to have the two reflections of the capillary on the same optical plane. A 3D reconstruction of the trajectories, captured with a single camera, is carried out along the vertical axis with a micrometer-scale precision. To investigate the interaction between the shear, the role of the gravity field, and motile microorganism, we track a model puller-type microalgae, \emph{Chlamydomonas reinhardtii} under a Poiseuille flow, using first its natural fluorescence and then a bright-field imaging. Understanding how confinement influences motility is crucial, and we show that this 3D tracking setup enables a full description of interactions between a motile organism and a solid border.

\end{abstract}

\maketitle

\section{\label{sec:level1}Introduction}

To ensure their survival, many microorganisms such as worms, fungi, bacteria and microalgae, have developed motility mechanisms that enable them to invade various environments. 
A number of studies are currently examining the transport mechanisms of these active particles.\cite{Lauga2020} In an unconfined medium, diffusion takes place with highly accelerated dynamics, compared with Brownian motion, thanks to self-propulsion mechanisms. Conversely, the swimming mechanisms can slow down the transport of microswimmers through a porous medium such as a foam.\cite{Roveillo2020} In this case, complex mechanisms such as the bouncing effects of microswimmers on the internal walls of pores could contribute to the blocking.\cite{thery2021rebound}
In the case of bacteria, and more extensively of \textit{pusher} microswimmers, the hydrodynamic interaction with the walls is attractive and plays a crucial role in the transport of microorganisms in a confined environment.\cite{Dentz2022} This leads the microswimmers to spend a significant portion of their trajectories exploring the walls instead of the bulk.\cite{Lauga2020, Berke2008, Baillou2023} In addition, the trajectory of elongated microswimmers in an inhomogeneous flow interacts strongly with the shear, an effect known as \textit{rheotaxis}, which can cause them to move upstream in the presence of a flow.\cite{Figueroa-Morales2015, Mathijssen2019} In the case of isotropic \textit{pullers} such as the flagellated microalgae \textit{Chlamydomonas reirhardtii} (CR), there is no  hydrodynamic attraction with boundary as in the case of \textit{pushers}; however swimming under confinement in a shear flow is also expected to lead to complex trajectories that could contribute explaining the macroscopic transport effects observed.\cite{Zottl2012} 

The study of all these behaviors requires visualization techniques. Standard microscopy has been widely used, using two-dimensional chambers with thicknesses ranging from 2 to 20 times the size of the organism to follow their two-dimensional (2D) trajectory.\cite{Ostapenko2018, thery2021rebound, Rafai2010, Volpe2011, Wu2000}
However, such confinement could have an effect on the microswimmer itself, due to the presence of walls at the top and bottom of the chamber. The microorganisms interact with the walls via various mechanisms: adhesion \cite{Kreis2019}, bouncing \cite{kanstler2009, contino2015, thery2021rebound}, hydrodynamic interactions leading to surface accumulation.\cite{Berke2008, Junot2022, Elgeti2015, buchner2021} These interactions with the boundary may affect the trajectory of the microswimmer in the plane of the chamber. Thanks to recent three-dimensional (3D) tracking techniques, it has been shown that the more bacteria are confined, the less they explore space laterally.\cite{Baillou2023} 3D tracking techniques is also particularly relevant for studying swimming trajectories of microswimmers in a shear flow, where the interaction between the shear flow and the motile particle lead to complex trajectories such as Bretherton-Jeffery trajectories.\cite{Junot2019}

Hence, 3D live imaging and tracking is needed to understand, at the level of single organisms, the interaction of the swimming motion with a shear flow in a confined environment. The last decade has seen a very important increase of 3D dynamic techniques to meet this need.\cite{bondoc2023methods} Five major 3D tracking techniques have been developed. 
(i)~In the \textit{General Defocusing Particle Tracking (GDPT)} technique, the patterns observed under the microscope when the object leaves its plane of focus is directly used to determine the depth position of the particles present in a single image, after a calibration using a reference set of images generated by particles of similar shapes.\cite{barnkob2020general} 
(ii)~In \textit{Digital Holographic Microscopy (DHM)}, a beamsplitter separates the incident beamlight of the microscope into a reference beam and a beam passing through the sample. After the transmission, both light beams are recombined to form a hologram on the surface of a camera's optical sensor: the hologram contains information not only on the position of the objects projected in the image plane, but also on their phase, which after computer processing enables depth to be determined.\cite{Bachimanchi2022, Memmolo2015} 
(iii)~Other 3D particle tracking techniques, such as the \textit{Stereo Darkfield Interferometry (SDI)}, involve customizing the microscope objective in order to split the image of the particles in two two images, with a relative distance depending on the particle depth along the optical axis. The separation of the images is achieved by inserting a wedge prism or glass slides.\cite{Yajima2008, Marumo2021, Rieu2021} This method enables displacements to be measured at the camera's acquisition frequency, with excellent resolution down to the sub-nanometric scale, depending on the magnification of the lens used. 
(iv)~The \textit{Lagrangian 3D tracking} device uses a motorized stage and a feedback loop to maintain the tracked particle on focus and in the centre of the image.\cite{Darnige2017, Junot2022} A real-time image processing determines the displacement of the stage to keep the chosen object at a fixed position in the observation frame, while the perpendicular displacement is based on the refocusing of the object to keep the moving object in focus. 

Unlike Eulerian tracking techniques, where visualization takes place in the laboratory reference frame, Lagrangian tracking allows the moving particle to be tracked over long periods, since the total tracking time is not limited by the particle leaving the field of view. However, it is limited to tracking a single particle, whereas Eulerian techniques can track all particles moving in the field of view. 
(v)~Another class of 3D tracking devices uses \textit{stereoscopic imaging} technique, which consists in visualizing at least two images of the object with two different viewing angles and reconstructing the 3D images by combining both images. Originally developed to study turbulent flows using seeded tracers in the fluid \cite{maas1993}, stereoscopic techniques are now used for 3D particle tracking velocimetry (PTV) and 3D Particle Image Velocimetry,\cite{Prasad2000, kim2011} as well as individual protist 3D tracking.\cite{Drescher2009} These systems conventionally require as many video cameras as there are angles of view to be acquired, which can make the system costly and electronically complex due to the need to synchronise all the cameras. To overcome these difficulties, some authors have developed custom-built optical devices that use ingenious assemblies of optical prisms and mirrors to obtain images from several different angles of view of the same object on the sensor of a single camera, making it possible to measure 3D flow velocity fields using a single-camera stereoscopy system.\cite{Cheung2005, Peterson2012} These technical simplifications have made it possible to produce low-cost and versatile tracking devices that can be carried out on board for field measurements in the ocean environment.\cite{Lertvilai2021}

In this article, we present a new laboratory device in the family of \textit{single camera stereoscopy} systems, designed to track planktonic particles in 3D, confined in a microchannel and subjected to a controlled low shear flow. The long-term objective of such an experimental scenario is to reproduce the situation of cell transport in a liquid foam's internal channels subjected to gravity drainage. The device allows to use the autofluorescence of the chlorophyll, in the case of phytoplankton, which enables very good signal-to-noise ratios to be achieved for the images, and the field of view to be widened so that particles can be tracked over fairly long periods of time. The custom-made device allows to visualize particles in a vertical flow, which is required in the case of planktonic particles sensitive to gravity. Moreover, the single-camera setup avoids the spatial and temporal synchronization problems inherent with the conventional multiple camera stereoscopy setup.

\section{Setup : Single camera stereoscopy with two mirrors}

\subsection{Experimental apparatus}

\begin{figure*}
\includegraphics[scale = 0.35]{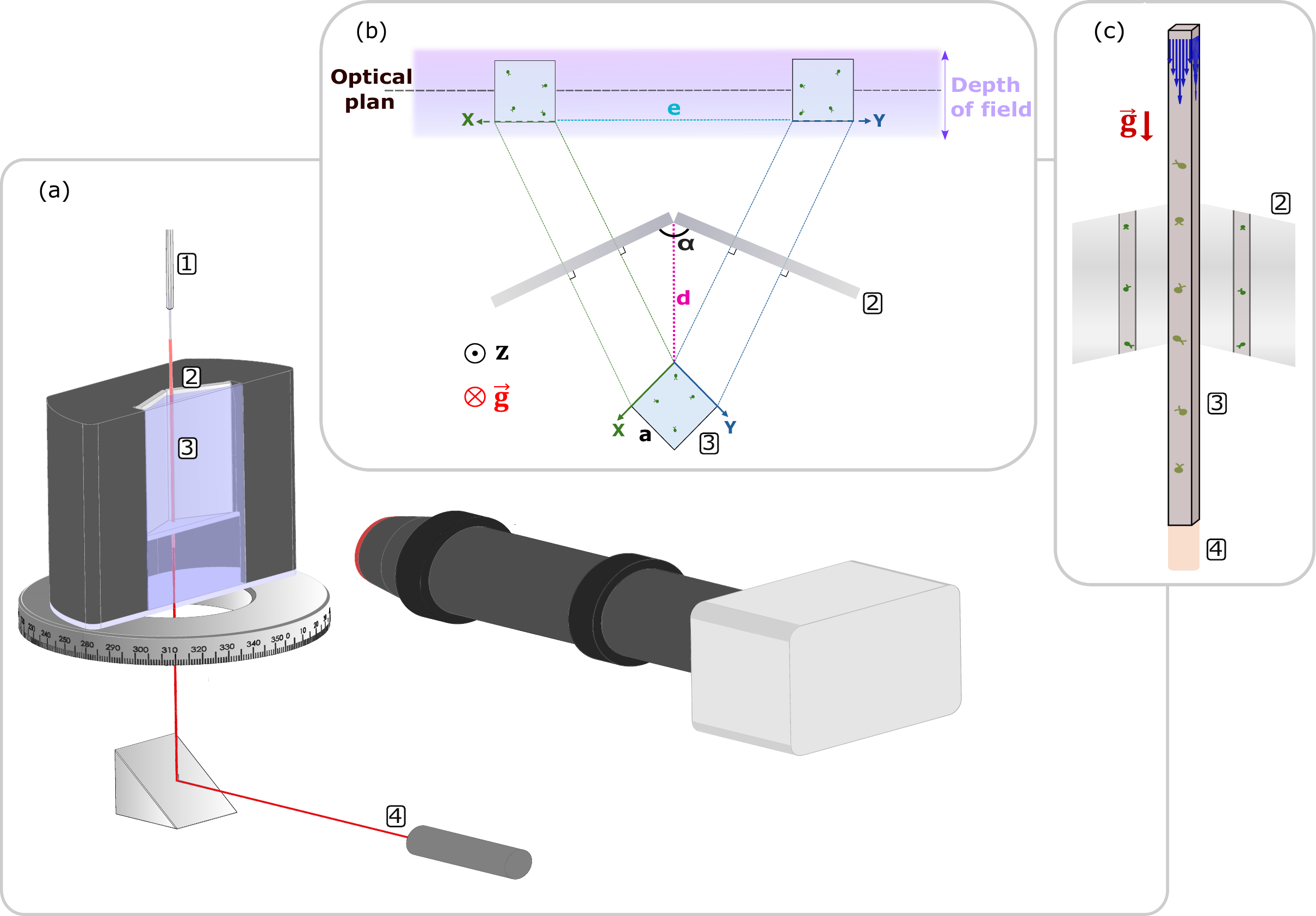}
\caption{\label{fig:setup}
(a) Schematic diagram of the experimental setup. The suspension of protists is injected through a needle connected to a syringe pump (not drawn here) in a vertical glass capillary having a square cross-section. The capillary is placed in front of two vertical mirrors forming an angle of $\alpha = 3\pi/4$. The phytoplanktonic cells are illuminated by a laser beam passing through the capillary and directed by a mirror placed under the capillary. The images of the light emitted by the cells in the capillary through both mirrors are visualized by a camera equipped with a zoom objective and recorded in a computer. In front of the camera objective, a band-pass filter is placed in order to select the fluorescence emission of the chlorophyll contained in the phytoplanktonic cells. The mirrors and capillary are placed in a chamber filled with water. The chamber contains in its lower part a horizontal plate pierced to allow the end of the capillary to pass through. Under the perforated plate, the end of the capillary is immersed in the water. The tank is placed on a rotating platform to adjust its orientation around the vertical axis. (b) Diagram of the square capillary and of its image through the two mirrors, seen from above. The $3\pi/4$ angle formed by the mirrors allows the image of the two perpendicular faces $(x, z)$ and $(y, z)$ of the capillary to be formed in the same plane. The optical plane on the camera and the depth of field are schematised in shaded violet. (c) Diagram of the side view of the square capillary illuminated by the laser beam and of its two images reflected by each of the two mirrors.\\
Legend: \fbox{1} Needle connected to syringe pump, \fbox{2} Mirrors, \fbox{3} Glass capillary, \fbox{4} Laser beam.
}
\end{figure*}

The single-camera stereoscopy setup is represented in Fig.~\ref{fig:setup}(a). It is based on the simultaneous visualization of the images of two perpendicular side views of the object of interest, achieved by imaging the object through two mirrors. At a macroscopic scale, this technique is generally used by placing the object in the field of view of the camera, and a mirror at 45~degrees to the side of the object, so as to obtain a front view and a side view of the object in the same image. However, the side view consists of an image of the object through the mirror, which is formed at a distance from the lens equal to the distance from the object plus twice the distance between the object and the mirror. To obtain a focused image of both the front and side views, the lens used must therefore have a depth of field at least equal to the distance between the object and its image in the mirror. Although this property is standard in optical camera lenses, microscope lenses have a much shallower depth of field, which means that two images at different distances from the lens cannot be simultaneously viewed on focus. Instead, the technique developed here involves using two mirrors, carefully oriented in relation to each other and to the sample so as to form two images of two perpendicular views in the same optical plane. This is achieved by forming an angle of $\alpha = 3\pi/4$ between the two mirrors, as schematized in figure ~\ref{fig:setup}b: the reflected images of the two perpendicular planes $(x, z)$ and $(y, z)$, placed symmetrically in relation to the bisector plane of the two mirrors, are aligned along the same plane behind the mirrors. Hence, the two reflections are located in the same object plane of the camera lens.

The sample used here is a hollow square glass capillary ($a$~=~200~µm square on the inside, 100~µm thick and 10~cm long), placed vertically and filled with a suspension of the unicellular phytoplanktonic algae \textit{Chlamydomonas reinhardtii} (CR). We use the autofluorescence of the chlorophyll within the cell to perform fluorescence imaging so as to select only the particles of interest in the images and improve the signal-to-noise ratio. \cite{finney2022} The suspension is illuminated by a laser beam (Laser Coherent Stingray, 10~mW) of wavelength 658~$\pm$~6~nm placed under the capillary. The capillary acts as a light waveguide for the laser beam, hence the cells are illuminated uniformly along the entire length of the channel. The sample is viewed using a camera (Hamamatsu Orca Flash~4.0) fitted with a Zoom Navitar lens (Zoom~6000 with 12~mm Focus and short 2X output adapter). A long-pass edge filter with a cutoff wavelength of 664~nm (Semrock RazorEdge® ultrasteep long-pass edge filter), placed in front of the lens, is used to isolate only the light emitted by the chloroplasts of the cells. 

Two vertical mirrors are placed behind the capillary. They are fixed to a 3D printed support so that they meet at their edges forming an angle of $\alpha = 3\pi/4$. In order to eliminate any parasitic optical reflection on the walls of the capillary of the fluorescent signal emitted by the cells, the mirrors and capillary are immersed in water to lower the optical index difference between the glass capillary and the outside. Furthermore, immersion in water enables the sample to be thermalized. The water tank is contained in a 3D printed plastic chamber, and a glass slide is glued to the front of the chamber to view the inside. In the lower part of the chamber (under the viewing window), a plate is pierced with a fine hole allowing the bottom of the capillary to pass through. In this way, when the cell suspension flows downwards into the capillary, the majority of the cells that leave the capillary through the bottom hole remain in the lower part of the chamber and do not rise into the viewing window.

To generate a controlled internal flow, the capillary is hermetically connected by a needle and tubes to a syringe fitted in a syringe pump (CETONI Nemesys S syringe pump) that enables stable continuous flow rates as low as tenths of microlitres per minute. Such value is required to realistically mimic the gravity-induced liquid flow within internal foam channels;\cite{Roveillo2020} furthermore, it results in a mean flow velocity comparable to that of the microswimmers, a regime yielding interesting effects in terms of interaction between the flow and their trajectories. \cite{Zottl2012}

The device has a number of elements for adjusting the correct alignment of all the parts of the assembly. Firstly, the capillary is held vertically at the top by a rotating platform (not shown on Fig.~\ref{fig:setup}a) that allows its orientation around the vertical axis to be adjusted. The correct orientation is obtained when one of the diagonals of the square section of the capillary is oriented parallel to the optical axis of the camera (Fig.~\ref{fig:setup}b), which can be set with the eye by optically aligning the two opposite vertical edges of the capillary. Secondly, the tank is placed on a rotating platform so that the mirrors can be oriented symmetrically in relation to the capillary (Fig.~\ref{fig:setup}a). This is also achieved with the eye, when the two mirror images of the two capillary faces are in the optical plane, i.e. the image of the perpendicular sides is no longer visible to the camera. Note that all these adjustments must be made under bright field illumination conditions, which can be achieved by temporarily removing the long pass edge filter at 664~nm placed in front of the camera lens. 

The whole device is placed in a black box to eliminate the effects of phototaxis due to external light. Organisms such as \textit{Chlamydomonas reinhardtii} exhibit phototactic responses to illumination with wavelengths below 650~nm;\cite{foster1980} they can therefore be observed with red light (such as the laser beam used here) without triggering any phototactic response.

The typical images acquired by the camera are shown on Fig.~\ref{fig_algo}a: in the middle of the image, the capillary is visible and out of focus, since the focus is in the plane of the images of the capillary through the two mirrors. Thus, the two focused images of the capillary, inclined at an angle $\pi/2$ to each other as seen in Fig.~\ref{fig:setup}c, are visible on figure \ref{fig_algo}a on either side of the capillary. Spatially separating the capillary from its two reflections on the image requires $e > a \sqrt{2}$ (see the notations in Fig. \ref{fig:setup}b). Geometrically, $e~=~2~d~\sin(\alpha)$, where $d$ is the shortest distance between the capillary and the edge between the two mirrors. Hence, once must have $d > a / (\sqrt{2} \sin\alpha)$, that is $d > a$ since $\alpha=3\pi/4$. Note that if necessary, the minimum distance $d$ can be reduced to $a / \sqrt{2}$ by reducing the angle $\alpha$ to $\pi / 2$; in this case the images of the capillary by the mirrors will no longer be images of two perpendicular planes, and the analysis of the images presented below to reconstruct the 3D image will have to be corrected. In the experiments presented here, we use $\alpha = 3 \pi / 4$ and $d$ = 1 mm.

The zoom lens allows the magnification to be adjusted. In order to track the microswimmers over the longest possible trajectories, we seek to obtain the widest possible field of vision, compatible however with the visualization of individual algae. This was achieved by setting a field of view of 6~mm~$\times$~6~mm, with a resolution of 2.9~µm~per~pixel. For this magnification, the depth of field of the zoom lens is larger that the capillary diameter, hence all the microswimmers captured in the images are in focus.

This setup allows to track microswimmer during up to 3 minutes without flow, and up to 80 seconds under a downwards flow rate of $0.3~\text{µL.min}^{-1}$.

\subsection{Tracking algorithm and 3D reconstruction}

\begin{figure}
\includegraphics[scale = 0.25]{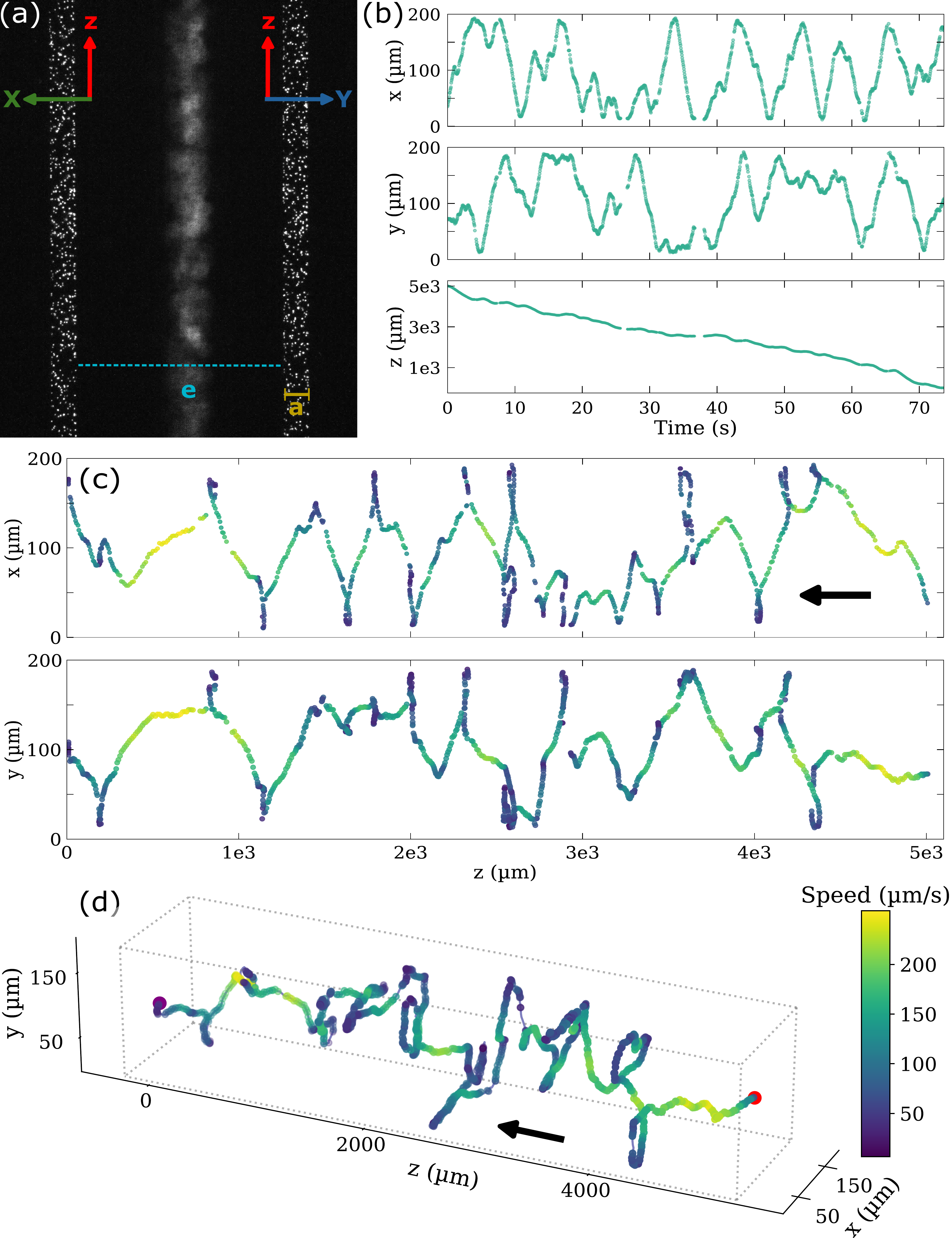}
\caption{\label{fig_algo} (a) Raw image in the plane containing the two reflections of the capillary produced by the mirrors (the blurred capillary is visible in the middle). Here the density of microalgae is $\sim$ 100 times higher than the one used during experiments. (b) Trajectories of an algae $x(t)$, $y(t)$ and $z(t)$ under a downwards flow (black arrow) $Q~=~0.3~\text{µL.min}^{-1}$. (c) Trajectories in the plane $(x,~z)$ and $(y,~z)$ of the algae; the black arrow indicate the direction of the flow. (d) 3D trajectory with $v(t)~=~\sqrt{v_x(t)^2~+~v_y(t)^2~+~v_z(t)^2}$ as a color. In panels (b), (c) and (d) as well as in panel (a), the gravity field is oriented in the direction of decreasing $z$.}
\end{figure}
To obtain 3D trajectories, 2D tracks have been performed on each of both images reflected by the mirrors that is either the (x, z) plane or the (y, z) plane (Fig.~\ref{fig_algo}c). A Python code using Trackpy, a freely available particle tracking package,\cite{trackpy} was employed. This package is based on an implementation of a version of the Crocker and Grier algorithm,\cite{crocker1996methods} which facilitates particle detection and the analysis of their trajectories.

Treatment of frames is necessary to improve the tracking quality: a bandpass filter was applied to the images, removing both low-frequency background variations and high-frequency noise. In addition, frames were divided into two parts to separate trajectories of the two reflections. Since the capillary is positioned vertically from the needle (where the flow is input) into the tank's hole, it may be slightly tilted. To correct this, the border positions are identified, and a fit is performed to find a correction matrix.

\begin{figure*}
\includegraphics[scale = 0.4]{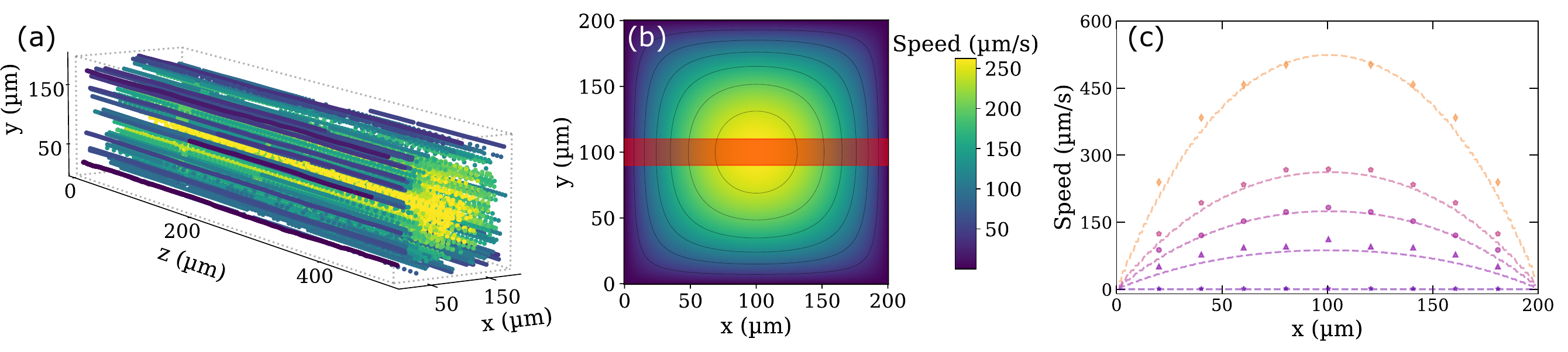}
\caption{\label{fig_poiseuille} (a) 3D trajectories of fluorescent beads (diameter~5~µm), submitted to a Poiseuille flow in the capillary, at a flow rate of $Q~=~0.3~\text{µL.min}^{-1}$; (b) Computed velocity profile $u(x, y)$ in the capillary cross-section, obtained from equation (\ref{uz}), for the same flow rate; (c) velocity profiles $v(x)$ for flow rates $ Q~=~0,~0.1,~0.2,~0.3,~0.6~\text{µL.min}^{-1}$ (from purple to yellow) obtained by tracking the beads within a thin layer 90~µm < $y$ < 110~µm represented in red in panel (b) and averaged along the $z$ direction; dashed line are obtained using equation (\ref{uz}) with no fitting parameter.}
\end{figure*}

The identification of individual positions in each frame was performed, considering parameters such as the size of the microswimmers, brightness, and background color. Next, these positions were linked to form trajectories, where the microswimmers's velocity in pixels per frame and the possible disappearance time were required. These latter two parameters are the most variable, depending on the flow and density of microswimmers in the experiments.

Subsequently, the vertical positions from both 2D tracks were recombined, and the differences between these vertical positions for all trajectories were computed. Multiple individuals may occupy the same vertical position, and frequent crossings between individuals are observed. To facilitate long-term tracking, a search for the minimal difference (below a specified threshold) along the vertical axis was conducted for each frame. When a match was found between the two tracks, the most recent horizontal position from an earlier frame was compared to ensure accurate trajectory reconstruction. This approach ensures that even if the algorithm mistakenly swaps the identity of an organism after a crossing while linking positions to form trajectories, the correct pair of positions is maintained. The two 2D tracks in Fig.~\ref{fig_algo}c are then recombined into a 3D trajectory, as illustrated in Fig.~\ref{fig_algo}d.

\section{Tracking performances}

\subsection{Poiseuille flow}
Since our system employs Eulerian tracking - that is the tracking of swimmer in the laboratory frame of reference - it can simultaneously track multiple microswimmers. To differentiate between the speed of the individual microswimmers and the flow speed, we simulate Poiseuille flow in a capillary with a square cross-section.\cite{delplace2018laminar} The origin of the coordinate system is set at the centre of the capillary, with $a$ representing the side length of the square cross-section. For a Newtonian fluid, the flow is parallel to the $z$ axis of the capillary, and the velocity profile is described by the following equation:

\begin{equation}
    \footnotesize u_z(x, y)=\frac{\pi Q}{2a^2} \frac{\sum^{+\infty}_{n = 1, 3, 5,...}\frac{(-1)^{(n-1)/2}}{n^3}\left(1-\frac{\cosh\left(\frac{n\pi}{a}y\right)}{\cosh\left(\frac{n\pi}{2}\right)}\right)\cos\left(\frac{n\pi}{a}x\right)}{\sum^{+\infty}_{n = 1, 3, 5,...} \frac{1}{n^4}\left(1-\frac{2}{\pi n} \tanh\left(\frac{n\pi}{2}\right)\right)}
    \label{uz}
\end{equation}

where $Q$ is the liquid flow rate. Thanks to this equation the flow velocity component experienced by a microswimmer based on its coordinates $(x,~y)$ within the capillary is calculated (Fig.~\ref{fig_poiseuille}b).

To validate the flow profile and conduct a test experiment, fluorescent beads (5~µm diameter, Firefli™ Fluorescent Green 468/508~nm) are introduced into the capillary as non-motile tracers. A laser with a wavelength of 405~nm is used to illuminate the beads, following the same setup as illustrated in Fig~\ref{fig:setup}. 

Fig.~\ref{fig_poiseuille}a shows that, as expected for passive microbeads in a Poiseuille flow, the bead trajectories remain parallel to the $z$ axis and the velocities are stationary. Hence we use the passive microbeads as tracers to measure the flow velocity profile: the comparison with equation (\ref{uz}) is very accurately obtained without any fitting parameter, as shown in figure~\ref{fig_poiseuille}c. This measurement shows that no convection occurs within the capillary, either under flow or at rest.

\subsection{Rebounds of Chlamydomonas reinhardtii against a solid wall under shear}

\begin{figure*}
    \includegraphics[scale = 0.268]{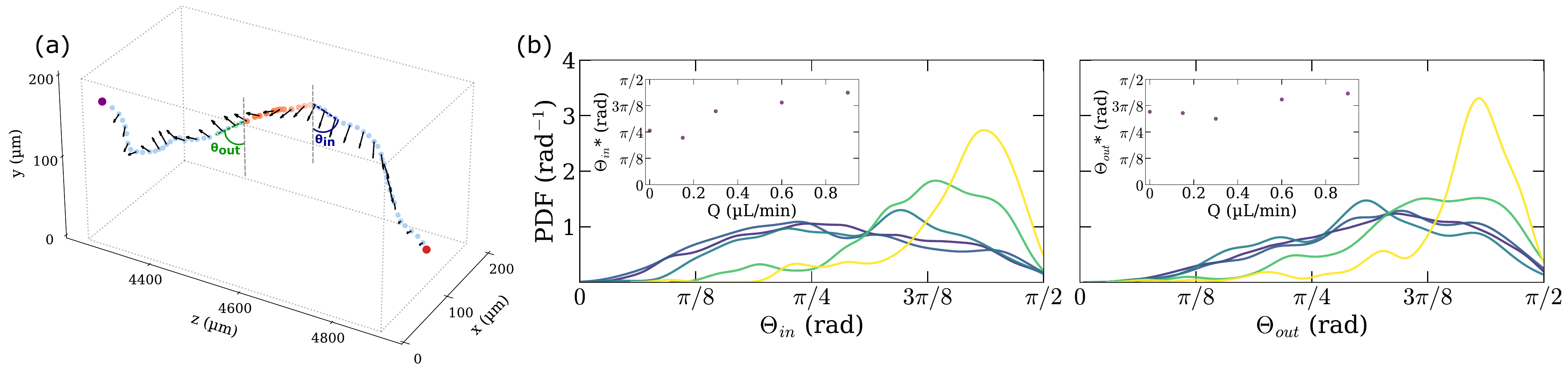} 
    \caption{\label{fig_rebound}~(a) 3D track of a CR algae scattered by the solid wall of the capillary, under a Poiseuille flow (flow rate $Q~=~0.3~\text{µL.min}^{-1}$). The microswimmer's intrinsic velocity, $v_{int}$, represented by the black arrows, is obtained by removing, at each point, the Poiseuille flow velocity (Eq. \ref{uz}) from the individual's velocity measured in the laboratory frame. Such a track allows to measure the angles $\theta_{in}$ and $\theta_{out}$ involved in the interaction with the wall. The portion of the trajectory within the region of thickness 20~µm from the boundary, that is defined as the region of interaction of the CR cell with the boundary, appears in orange; the angle $\theta_{in}$ (resp. $\theta_{out}$) is defined when the cell is entering (resp. exiting) this region. (b) PDF of the angles $\theta_{in}$ and $\theta_{out}$ at the boundary, for various flow rates, $Q~=~0, 0.15, 0.3, 0.6, 0.9~\text{µL.min}^{-1}$ color-coded from dark blue to yellow. Inserts: mode of the PDF as a function of the liquid flow rate.}
\end{figure*}

As an illustration, we show that our setup enables tracking the trajectory of the puller biflagellated microswimmer CR against a rigid boundary. The re-orientation of the algae velocity can then be recorded and analysed. Such data are of great value to understand the interaction of the CR with a boundary and how this interaction is affected by an external flow, an ingredient that is required in order to fully understand the transport of CR motile algae through a porous medium.

We used the strain CC124$-$ of CR. The algae were kept on High Salt with Acetate (HSA) medium agar plate. For experiments, algae were propagated in liquid HSA medium on an orbital shaker in an incubator at 23$^{\circ}~$C on a 12h/12h bright/dark light cycle. Cells were used between 48h and 72h after inoculation in liquid medium. The cell concentration in HSA was adapted in order to reach a cell concentration of the order of 10$^8$~cells.L$^{-1}$, a fairly diluted situation in which the distance between cells is of the order of 50~times the cell diameter, close to the expected situation met in natural marine foams. \cite{Roveillo2020} The cell suspension is placed in a glass syringe and carefully injected into the capillary within the single camera spectroscopy setup, at a prescribed flow rate, to enable 3D tracking of microswimmers.

The instantaneous velocity $\vec{v}(x,y,z)$ of the cell along its trajectory is a combination of the Poiseuille flow velocity $\vec{u}(x, y)~=~u_z(x,~y)~\vec{e}_z$ given in equation (\ref{uz}) and of the swimming cell intrinsic velocity $\vec{v}_{int}(x, y, z)$. In Fig. \ref{fig_rebound}a, the intrinsic velocity of the microswimmer is shown by arrows along the cell trajectory. 

The trajectory tracking while the motile cell is meeting a boundary allows to define bouncing angles $\theta_{in}$ and $\theta_{out}$ that are formed between the tangent of the microalgae trajectory and the normal to the boundary just before and just after the bounce. In Fig.~\ref{fig_rebound}a, we chose to define the boundary region as a layer having an extension of 20~µm from the boundary interface $y~=~$200~µm, since 20~µm is the typical length of a CR cell, including its flagella. We define the tangent to the trajectory by averaging the orientation of the trajectory during a time lapse of 0.25~s. Such definitions meet the angle determination performed in several other studies from the literature. \cite{kanstler2009, thery2021rebound, contino2015, buchner2021} As a result, Fig.~\ref{fig_rebound}b shows the Probability Distribution Function (PDF) of the angles $\theta_{in}$ and $\theta_{out}$, for different flow rate intensities in the capillary. The data was acquired considering approximately 200 rebounds per $Q$ value, going to 1800 values without flow. While the angles are rather equiprobable for low flow rate, the probability distribution becomes pitted at large flow rates; the most probable values of the impact angles $\theta_{in}$ and $\theta_{out}$ increase with the flow rate to approach $\pi/2$, which corresponds to grazing incidence, i.e. a trajectory aligned with the flow.
Surprisingly, this change in PDF occurs abruptly when the flow rate becomes greater than 0.3~µL.min$^{-1}$, which corresponds to an average flow velocity of $<u>$~=~125~µm.s$^{-1}$, a value comparable to the average intrinsic swimming velocity of the CR cells.

\section{Outlooks/versatility}

\subsection{Bright field illumination}

\begin{figure}[h!]
    \includegraphics[scale = 0.35]{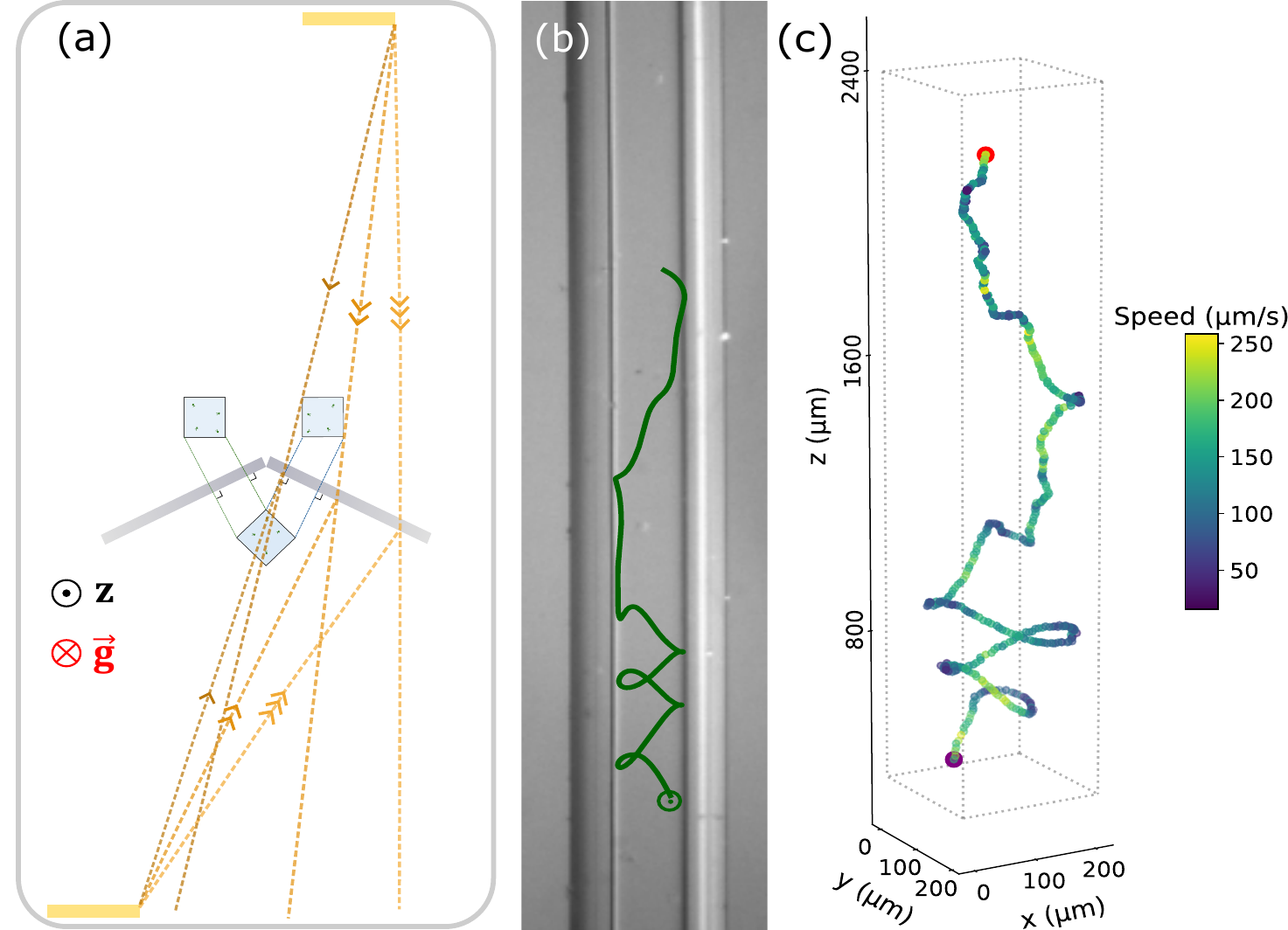}
    \caption{\label{fig_bright_field}~(a) Sketch of the adapted setup for a bright field illumination. The square capillary and its image through the two mirrors are represented from above as in Fig. \ref{fig:setup}b. In addition, the setup is illuminated using a light beam placed beneath the camera and tilted from about 20 degrees from the horizontal. Two stripes of white paper are placed (one of them is sketched in yellow on the diagram) on either side of the camera. (b) Raw image in the plane of the two images of the capillary by the mirrors in bright field illumination. A CR microswimmer appears as a dark point ; the CR trajectory is sketched in green on the images. The flow rate is $Q~=~0.3~\text{µL.min}^{-1}$ in the capillary. (c) 3D reconstructed trajectory.}
\end{figure}

Not all microswimmers are naturally fluorescent; however, the setup presented in this article can be customised in order to be used in bright field conditions. To do this, the laser lightning (Fig.~\ref{fig:setup}a) is removed, and a beam of collimated light placed beneath the camera illuminates the device from the front. The plane of incidence is defined by the $z$-axis and the optical axis of the camera objective, with the beam directed at about a 20$^{\circ}$ angle to the horizontal to avoid reflected beams within the camera’s field of view. The beam diameter is 3~cm, and the illumination is a white light, filtered using a high pass filter to eliminate wavelengths below 630~nm. Two strips cut from a diffusing material (typically white paper) are placed in the path of the incident light on either side of the camera. They are shown in yellow in Fig.~\ref{fig_bright_field}a, which represents a top view of the device. In this way, the light scattered isotropically by one of the thin strips is reflected on the opposite mirrors towards the camera objective, imitating illumination from behind the capillary. The microswimmers present in the capillary then appear in grey on a white background, just as if a bright field device were being used. The contrast is then sufficient to allow the microswimmers to be tracked, as shown in figure \ref{fig_bright_field}b.

\subsection{Transform any microscope into a 3D particle tracking setup.}
The principle of the setup using two mirrors making an angle of $3\pi /4$ is very versatile and can be adapted on any binocular magnifier or microscope fitted with a long distance objective. This is achieved by 3D printing a support piece onto which mirrors can be glued at the appropriate angle. The capillary to be observed is then placed between this part and the objective. Depending on the organism to be imaged, the microscope can be illuminated either in bright field or by fluorescence. One can then focus on the two images of the capillary through the mirrors and follow the 3D movement of the particles.

\section{Conclusion}

We presented a 3D tracking technique to follow particles under flow using few accessible tools, such as capillaries, two mirrors and a camera. The simplicity of this system enables an easy reproductibility and a combination with other devices, like microscopes. With this stereoscopic method, it is possible to follow individuals with an Eulerian point of view for duration going for several minutes without flow to superior to several tens of seconds for a liquid flow rate of $0.9~\text{µL.min}^{-1}$ in the capillary. Since we are only limited by the surface resolution of mirrors, particles can be from the micron size to hundreds of microns and the medium accepts various geometries.

In this article, we show that in order to investigate the role of the shear flow rate in the scattering angle on the solid boundaries, one can deduce the speed of microswimmers inside a flow, by removing the flow speed calculated with the position of the individual. Further studies are needed to complete these analyses, and we believe that our setup might help achieve a better understanding of interaction between the swimming trajectory and the boundary, an element that is still lacking to build theoretical models to describe the transport of puller microswimmers in porous media.

\begin{acknowledgments}
We wish to thank many researchers and engineers who helped in all the different aspects of this work, and for fruitful discussions: Thierry Savy, Olivier Cardoso, Eric Clément, Thierry Darnige, Albane Théry, Amaury Fourgeaud. This work was financially supported by the IdEx Université Paris Cité, ANR-18-IDEX-0001 and by the Agence Nationale de la Recherche (ANR) through the project ECUME, ANR-22-06CE-0029 and the project PUSHPULL, ANR-22-CE30-0038.
\end{acknowledgments}

\bibliography{Bibliographie}

\begin{thebibliography}{40}%
\makeatletter
\providecommand \@ifxundefined [1]{%
 \@ifx{#1\undefined}
}%
\providecommand \@ifnum [1]{%
 \ifnum #1\expandafter \@firstoftwo
 \else \expandafter \@secondoftwo
 \fi
}%
\providecommand \@ifx [1]{%
 \ifx #1\expandafter \@firstoftwo
 \else \expandafter \@secondoftwo
 \fi
}%
\providecommand \natexlab [1]{#1}%
\providecommand \enquote  [1]{``#1''}%
\providecommand \bibnamefont  [1]{#1}%
\providecommand \bibfnamefont [1]{#1}%
\providecommand \citenamefont [1]{#1}%
\providecommand \href@noop [0]{\@secondoftwo}%
\providecommand \href [0]{\begingroup \@sanitize@url \@href}%
\providecommand \@href[1]{\@@startlink{#1}\@@href}%
\providecommand \@@href[1]{\endgroup#1\@@endlink}%
\providecommand \@sanitize@url [0]{\catcode `\\12\catcode `\$12\catcode
  `\&12\catcode `\#12\catcode `\^12\catcode `\_12\catcode `\%12\relax}%
\providecommand \@@startlink[1]{}%
\providecommand \@@endlink[0]{}%
\providecommand \url  [0]{\begingroup\@sanitize@url \@url }%
\providecommand \@url [1]{\endgroup\@href {#1}{\urlprefix }}%
\providecommand \urlprefix  [0]{URL }%
\providecommand \Eprint [0]{\href }%
\providecommand \doibase [0]{http://dx.doi.org/}%
\providecommand \selectlanguage [0]{\@gobble}%
\providecommand \bibinfo  [0]{\@secondoftwo}%
\providecommand \bibfield  [0]{\@secondoftwo}%
\providecommand \translation [1]{[#1]}%
\providecommand \BibitemOpen [0]{}%
\providecommand \bibitemStop [0]{}%
\providecommand \bibitemNoStop [0]{.\EOS\space}%
\providecommand \EOS [0]{\spacefactor3000\relax}%
\providecommand \BibitemShut  [1]{\csname bibitem#1\endcsname}%
\let\auto@bib@innerbib\@empty
\bibitem [{\citenamefont {Lauga}(2020)}]{Lauga2020}%
  \BibitemOpen
  \bibfield  {author} {\bibinfo {author} {\bibfnamefont {E.}~\bibnamefont
  {Lauga}},\ }\href@noop {} {\emph {\bibinfo {title} {The Fluid Dynamics of
  Cell Motility}}}\ (\bibinfo  {publisher} {Cambridge University Press},\
  \bibinfo {address} {Cambridge},\ \bibinfo {year} {2020})\BibitemShut
  {NoStop}%
\bibitem [{\citenamefont {Roveillo}\ \emph {et~al.}(2020)\citenamefont
  {Roveillo}, \citenamefont {Dervaux}, \citenamefont {Wang}, \citenamefont
  {Rouyer}, \citenamefont {Zanchi}, \citenamefont {Seuront},\ and\
  \citenamefont {Elias}}]{Roveillo2020}%
  \BibitemOpen
  \bibfield  {author} {\bibinfo {author} {\bibfnamefont {Q.}~\bibnamefont
  {Roveillo}}, \bibinfo {author} {\bibfnamefont {J.}~\bibnamefont {Dervaux}},
  \bibinfo {author} {\bibfnamefont {Y.}~\bibnamefont {Wang}}, \bibinfo {author}
  {\bibfnamefont {F.}~\bibnamefont {Rouyer}}, \bibinfo {author} {\bibfnamefont
  {D.}~\bibnamefont {Zanchi}}, \bibinfo {author} {\bibfnamefont
  {L.}~\bibnamefont {Seuront}}, \ and\ \bibinfo {author} {\bibfnamefont
  {F.}~\bibnamefont {Elias}},\ }\href@noop {} {\bibfield  {journal} {\bibinfo
  {journal} {Journal of The Royal Society Interface}\ }\textbf {\bibinfo
  {volume} {17}},\ \bibinfo {pages} {20200077} (\bibinfo {year}
  {2020})}\BibitemShut {NoStop}%
\bibitem [{\citenamefont {Th{\'e}ry}\ \emph {et~al.}(2021)\citenamefont
  {Th{\'e}ry}, \citenamefont {Wang}, \citenamefont {Dvoriashyna}, \citenamefont
  {Eloy}, \citenamefont {Elias},\ and\ \citenamefont
  {Lauga}}]{thery2021rebound}%
  \BibitemOpen
  \bibfield  {author} {\bibinfo {author} {\bibfnamefont {A.}~\bibnamefont
  {Th{\'e}ry}}, \bibinfo {author} {\bibfnamefont {Y.}~\bibnamefont {Wang}},
  \bibinfo {author} {\bibfnamefont {M.}~\bibnamefont {Dvoriashyna}}, \bibinfo
  {author} {\bibfnamefont {C.}~\bibnamefont {Eloy}}, \bibinfo {author}
  {\bibfnamefont {F.}~\bibnamefont {Elias}}, \ and\ \bibinfo {author}
  {\bibfnamefont {E.}~\bibnamefont {Lauga}},\ }\href@noop {} {\bibfield
  {journal} {\bibinfo  {journal} {Soft Matter}\ }\textbf {\bibinfo {volume}
  {17}},\ \bibinfo {pages} {4857} (\bibinfo {year} {2021})}\BibitemShut
  {NoStop}%
\bibitem [{\citenamefont {Dentz}\ \emph {et~al.}(2022)\citenamefont {Dentz},
  \citenamefont {Creppy}, \citenamefont {Douarche}, \citenamefont
  {Cl{\'e}ment},\ and\ \citenamefont {Auradou}}]{Dentz2022}%
  \BibitemOpen
  \bibfield  {author} {\bibinfo {author} {\bibfnamefont {M.}~\bibnamefont
  {Dentz}}, \bibinfo {author} {\bibfnamefont {A.}~\bibnamefont {Creppy}},
  \bibinfo {author} {\bibfnamefont {C.}~\bibnamefont {Douarche}}, \bibinfo
  {author} {\bibfnamefont {E.}~\bibnamefont {Cl{\'e}ment}}, \ and\ \bibinfo
  {author} {\bibfnamefont {H.}~\bibnamefont {Auradou}},\ }\href@noop {}
  {\bibfield  {journal} {\bibinfo  {journal} {Journal of Fluid Mechanics}\
  }\textbf {\bibinfo {volume} {946}},\ \bibinfo {pages} {A33} (\bibinfo {year}
  {2022})}\BibitemShut {NoStop}%
\bibitem [{\citenamefont {Berke}\ \emph {et~al.}(2008)\citenamefont {Berke},
  \citenamefont {Turner}, \citenamefont {Berg},\ and\ \citenamefont
  {Lauga}}]{Berke2008}%
  \BibitemOpen
  \bibfield  {author} {\bibinfo {author} {\bibfnamefont {A.~P.}\ \bibnamefont
  {Berke}}, \bibinfo {author} {\bibfnamefont {L.}~\bibnamefont {Turner}},
  \bibinfo {author} {\bibfnamefont {H.~C.}\ \bibnamefont {Berg}}, \ and\
  \bibinfo {author} {\bibfnamefont {E.}~\bibnamefont {Lauga}},\ }\href@noop {}
  {\bibfield  {journal} {\bibinfo  {journal} {Phys. Rev. Lett.}\ }\textbf
  {\bibinfo {volume} {101}},\ \bibinfo {pages} {038102} (\bibinfo {year}
  {2008})}\BibitemShut {NoStop}%
\bibitem [{\citenamefont {Baillou}(2023)}]{Baillou2023}%
  \BibitemOpen
  \bibfield  {author} {\bibinfo {author} {\bibfnamefont {R.}~\bibnamefont
  {Baillou}},\ }\emph {\bibinfo {title} {Exploration lagrangienne des
  environnements complexes par les micro-organismes: suivi Lagrangien de E.
  coli motiles sous confinement et p{\'e}n{\'e}tration de la barri{\`e}re de
  mucus}},\ \href@noop {} {Ph.D. thesis},\ \bibinfo  {school} {Sorbonne
  universit{\'e}} (\bibinfo {year} {2023})\BibitemShut {NoStop}%
\bibitem [{\citenamefont {Figueroa-Morales}\ \emph {et~al.}(2015)\citenamefont
  {Figueroa-Morales}, \citenamefont {Leonardo~Mi{\~n}o}, \citenamefont
  {Rivera}, \citenamefont {Caballero}, \citenamefont {Cl{\'e}ment},
  \citenamefont {Altshuler},\ and\ \citenamefont
  {Lindner}}]{Figueroa-Morales2015}%
  \BibitemOpen
  \bibfield  {author} {\bibinfo {author} {\bibfnamefont {N.}~\bibnamefont
  {Figueroa-Morales}}, \bibinfo {author} {\bibfnamefont {G.}~\bibnamefont
  {Leonardo~Mi{\~n}o}}, \bibinfo {author} {\bibfnamefont {A.}~\bibnamefont
  {Rivera}}, \bibinfo {author} {\bibfnamefont {R.}~\bibnamefont {Caballero}},
  \bibinfo {author} {\bibfnamefont {E.}~\bibnamefont {Cl{\'e}ment}}, \bibinfo
  {author} {\bibfnamefont {E.}~\bibnamefont {Altshuler}}, \ and\ \bibinfo
  {author} {\bibfnamefont {A.}~\bibnamefont {Lindner}},\ }\href@noop {}
  {\bibfield  {journal} {\bibinfo  {journal} {Soft Matter}\ }\textbf {\bibinfo
  {volume} {11}},\ \bibinfo {pages} {6284} (\bibinfo {year}
  {2015})}\BibitemShut {NoStop}%
\bibitem [{\citenamefont {Mathijssen}\ \emph {et~al.}(2019)\citenamefont
  {Mathijssen}, \citenamefont {Figueroa-Morales}, \citenamefont {Junot},
  \citenamefont {Cl{\'e}ment}, \citenamefont {Lindner},\ and\ \citenamefont
  {Z{\"o}ttl}}]{Mathijssen2019}%
  \BibitemOpen
  \bibfield  {author} {\bibinfo {author} {\bibfnamefont {A.~J. T.~M.}\
  \bibnamefont {Mathijssen}}, \bibinfo {author} {\bibfnamefont
  {N.}~\bibnamefont {Figueroa-Morales}}, \bibinfo {author} {\bibfnamefont
  {G.}~\bibnamefont {Junot}}, \bibinfo {author} {\bibfnamefont
  {{\'E}.}~\bibnamefont {Cl{\'e}ment}}, \bibinfo {author} {\bibfnamefont
  {A.}~\bibnamefont {Lindner}}, \ and\ \bibinfo {author} {\bibfnamefont
  {A.}~\bibnamefont {Z{\"o}ttl}},\ }\href@noop {} {\bibfield  {journal}
  {\bibinfo  {journal} {Nature Communications}\ }\textbf {\bibinfo {volume}
  {10}},\ \bibinfo {pages} {3434} (\bibinfo {year} {2019})}\BibitemShut
  {NoStop}%
\bibitem [{\citenamefont {Z{\"o}ttl}\ and\ \citenamefont
  {Stark}(2012)}]{Zottl2012}%
  \BibitemOpen
  \bibfield  {author} {\bibinfo {author} {\bibfnamefont {A.}~\bibnamefont
  {Z{\"o}ttl}}\ and\ \bibinfo {author} {\bibfnamefont {H.}~\bibnamefont
  {Stark}},\ }\href@noop {} {\bibfield  {journal} {\bibinfo  {journal}
  {Physical Review Letters}\ }\textbf {\bibinfo {volume} {108}},\ \bibinfo
  {pages} {218104} (\bibinfo {year} {2012})}\BibitemShut {NoStop}%
\bibitem [{\citenamefont {Ostapenko}\ \emph {et~al.}(2018)\citenamefont
  {Ostapenko}, \citenamefont {Schwarzendahl}, \citenamefont {B\"oddeker},
  \citenamefont {Kreis}, \citenamefont {Cammann}, \citenamefont {Mazza},\ and\
  \citenamefont {B\"aumchen}}]{Ostapenko2018}%
  \BibitemOpen
  \bibfield  {author} {\bibinfo {author} {\bibfnamefont {T.}~\bibnamefont
  {Ostapenko}}, \bibinfo {author} {\bibfnamefont {F.~J.}\ \bibnamefont
  {Schwarzendahl}}, \bibinfo {author} {\bibfnamefont {T.~J.}\ \bibnamefont
  {B\"oddeker}}, \bibinfo {author} {\bibfnamefont {C.~T.}\ \bibnamefont
  {Kreis}}, \bibinfo {author} {\bibfnamefont {J.}~\bibnamefont {Cammann}},
  \bibinfo {author} {\bibfnamefont {M.~G.}\ \bibnamefont {Mazza}}, \ and\
  \bibinfo {author} {\bibfnamefont {O.}~\bibnamefont {B\"aumchen}},\
  }\href@noop {} {\bibfield  {journal} {\bibinfo  {journal} {Phys. Rev. Lett.}\
  }\textbf {\bibinfo {volume} {120}},\ \bibinfo {pages} {068002} (\bibinfo
  {year} {2018})}\BibitemShut {NoStop}%
\bibitem [{\citenamefont {Rafa\"{\i}}\ \emph {et~al.}(2010)\citenamefont
  {Rafa\"{\i}}, \citenamefont {Jibuti},\ and\ \citenamefont
  {Peyla}}]{Rafai2010}%
  \BibitemOpen
  \bibfield  {author} {\bibinfo {author} {\bibfnamefont {S.}~\bibnamefont
  {Rafa\"{\i}}}, \bibinfo {author} {\bibfnamefont {L.}~\bibnamefont {Jibuti}},
  \ and\ \bibinfo {author} {\bibfnamefont {P.}~\bibnamefont {Peyla}},\
  }\href@noop {} {\bibfield  {journal} {\bibinfo  {journal} {Phys. Rev. Lett.}\
  }\textbf {\bibinfo {volume} {104}},\ \bibinfo {pages} {098102} (\bibinfo
  {year} {2010})}\BibitemShut {NoStop}%
\bibitem [{\citenamefont {Volpe}\ \emph {et~al.}(2011)\citenamefont {Volpe},
  \citenamefont {Buttinoni}, \citenamefont {Vogt}, \citenamefont
  {K{\"u}mmerer},\ and\ \citenamefont {Bechinger}}]{Volpe2011}%
  \BibitemOpen
  \bibfield  {author} {\bibinfo {author} {\bibfnamefont {G.}~\bibnamefont
  {Volpe}}, \bibinfo {author} {\bibfnamefont {I.}~\bibnamefont {Buttinoni}},
  \bibinfo {author} {\bibfnamefont {D.}~\bibnamefont {Vogt}}, \bibinfo {author}
  {\bibfnamefont {H.-J.}\ \bibnamefont {K{\"u}mmerer}}, \ and\ \bibinfo
  {author} {\bibfnamefont {C.}~\bibnamefont {Bechinger}},\ }\href@noop {}
  {\bibfield  {journal} {\bibinfo  {journal} {Soft Matter}\ }\textbf {\bibinfo
  {volume} {7}},\ \bibinfo {pages} {8810} (\bibinfo {year} {2011})}\BibitemShut
  {NoStop}%
\bibitem [{\citenamefont {Wu}\ and\ \citenamefont {Libchaber}(2000)}]{Wu2000}%
  \BibitemOpen
  \bibfield  {author} {\bibinfo {author} {\bibfnamefont {X.~L.}\ \bibnamefont
  {Wu}}\ and\ \bibinfo {author} {\bibfnamefont {A.}~\bibnamefont {Libchaber}},\
  }\href@noop {} {\bibfield  {journal} {\bibinfo  {journal} {Phys Rev Lett}\
  }\textbf {\bibinfo {volume} {84}},\ \bibinfo {pages} {3017} (\bibinfo {year}
  {2000})}\BibitemShut {NoStop}%
\bibitem [{\citenamefont {Kreis}\ \emph {et~al.}(2019)\citenamefont {Kreis},
  \citenamefont {Grangier},\ and\ \citenamefont {B{\"a}umchen}}]{Kreis2019}%
  \BibitemOpen
  \bibfield  {author} {\bibinfo {author} {\bibfnamefont {C.~T.}\ \bibnamefont
  {Kreis}}, \bibinfo {author} {\bibfnamefont {A.}~\bibnamefont {Grangier}}, \
  and\ \bibinfo {author} {\bibfnamefont {O.}~\bibnamefont {B{\"a}umchen}},\
  }\href@noop {} {\bibfield  {journal} {\bibinfo  {journal} {Soft Matter}\
  }\textbf {\bibinfo {volume} {15}},\ \bibinfo {pages} {3027} (\bibinfo {year}
  {2019})}\BibitemShut {NoStop}%
\bibitem [{\citenamefont {Kantsler}\ \emph {et~al.}(2013)\citenamefont
  {Kantsler}, \citenamefont {Dunkel}, \citenamefont {Polin},\ and\
  \citenamefont {Goldstein}}]{kanstler2009}%
  \BibitemOpen
  \bibfield  {author} {\bibinfo {author} {\bibfnamefont {V.}~\bibnamefont
  {Kantsler}}, \bibinfo {author} {\bibfnamefont {J.}~\bibnamefont {Dunkel}},
  \bibinfo {author} {\bibfnamefont {M.}~\bibnamefont {Polin}}, \ and\ \bibinfo
  {author} {\bibfnamefont {R.~E.}\ \bibnamefont {Goldstein}},\ }\href@noop {}
  {\bibfield  {journal} {\bibinfo  {journal} {Proceedings of the National
  Academy of Sciences}\ }\textbf {\bibinfo {volume} {110}},\ \bibinfo {pages}
  {1187} (\bibinfo {year} {2013})}\BibitemShut {NoStop}%
\bibitem [{\citenamefont {Contino}\ \emph {et~al.}(2015)\citenamefont
  {Contino}, \citenamefont {Lushi}, \citenamefont {Tuval}, \citenamefont
  {Kantsler},\ and\ \citenamefont {Polin}}]{contino2015}%
  \BibitemOpen
  \bibfield  {author} {\bibinfo {author} {\bibfnamefont {M.}~\bibnamefont
  {Contino}}, \bibinfo {author} {\bibfnamefont {E.}~\bibnamefont {Lushi}},
  \bibinfo {author} {\bibfnamefont {I.}~\bibnamefont {Tuval}}, \bibinfo
  {author} {\bibfnamefont {V.}~\bibnamefont {Kantsler}}, \ and\ \bibinfo
  {author} {\bibfnamefont {M.}~\bibnamefont {Polin}},\ }\href@noop {}
  {\bibfield  {journal} {\bibinfo  {journal} {Phys. Rev. Lett.}\ }\textbf
  {\bibinfo {volume} {115}},\ \bibinfo {pages} {258102} (\bibinfo {year}
  {2015})}\BibitemShut {NoStop}%
\bibitem [{\citenamefont {Junot}\ \emph {et~al.}(2022)\citenamefont {Junot},
  \citenamefont {Darnige}, \citenamefont {Lindner}, \citenamefont {Martinez},
  \citenamefont {Arlt}, \citenamefont {Dawson}, \citenamefont {Poon},
  \citenamefont {Auradou},\ and\ \citenamefont {Cl\'ement}}]{Junot2022}%
  \BibitemOpen
  \bibfield  {author} {\bibinfo {author} {\bibfnamefont {G.}~\bibnamefont
  {Junot}}, \bibinfo {author} {\bibfnamefont {T.}~\bibnamefont {Darnige}},
  \bibinfo {author} {\bibfnamefont {A.}~\bibnamefont {Lindner}}, \bibinfo
  {author} {\bibfnamefont {V.~A.}\ \bibnamefont {Martinez}}, \bibinfo {author}
  {\bibfnamefont {J.}~\bibnamefont {Arlt}}, \bibinfo {author} {\bibfnamefont
  {A.}~\bibnamefont {Dawson}}, \bibinfo {author} {\bibfnamefont {W.~C.~K.}\
  \bibnamefont {Poon}}, \bibinfo {author} {\bibfnamefont {H.}~\bibnamefont
  {Auradou}}, \ and\ \bibinfo {author} {\bibfnamefont {E.}~\bibnamefont
  {Cl\'ement}},\ }\href@noop {} {\bibfield  {journal} {\bibinfo  {journal}
  {Phys. Rev. Lett.}\ }\textbf {\bibinfo {volume} {128}},\ \bibinfo {pages}
  {248101} (\bibinfo {year} {2022})}\BibitemShut {NoStop}%
\bibitem [{\citenamefont {Elgeti}\ \emph {et~al.}(2015)\citenamefont {Elgeti},
  \citenamefont {Winkler},\ and\ \citenamefont {Gompper}}]{Elgeti2015}%
  \BibitemOpen
  \bibfield  {author} {\bibinfo {author} {\bibfnamefont {J.}~\bibnamefont
  {Elgeti}}, \bibinfo {author} {\bibfnamefont {R.~G.}\ \bibnamefont {Winkler}},
  \ and\ \bibinfo {author} {\bibfnamefont {G.}~\bibnamefont {Gompper}},\
  }\href@noop {} {\bibfield  {journal} {\bibinfo  {journal} {Reports on
  progress in physics}\ }\textbf {\bibinfo {volume} {78}},\ \bibinfo {pages}
  {056601} (\bibinfo {year} {2015})}\BibitemShut {NoStop}%
\bibitem [{\citenamefont {Buchner}\ \emph {et~al.}(2021)\citenamefont
  {Buchner}, \citenamefont {Muller}, \citenamefont {Mehmood},\ and\
  \citenamefont {Tam}}]{buchner2021}%
  \BibitemOpen
  \bibfield  {author} {\bibinfo {author} {\bibfnamefont {A.-J.}\ \bibnamefont
  {Buchner}}, \bibinfo {author} {\bibfnamefont {K.}~\bibnamefont {Muller}},
  \bibinfo {author} {\bibfnamefont {J.}~\bibnamefont {Mehmood}}, \ and\
  \bibinfo {author} {\bibfnamefont {D.}~\bibnamefont {Tam}},\ }\href {\doibase
  10.1073/pnas.2102095118} {\bibfield  {journal} {\bibinfo  {journal}
  {Proceedings of the National Academy of Sciences}\ }\textbf {\bibinfo
  {volume} {118}},\ \bibinfo {pages} {e2102095118} (\bibinfo {year} {2021})},\
  \Eprint
  {http://arxiv.org/abs/https://www.pnas.org/doi/pdf/10.1073/pnas.2102095118}
  {https://www.pnas.org/doi/pdf/10.1073/pnas.2102095118} \BibitemShut {NoStop}%
\bibitem [{\citenamefont {Junot}\ \emph {et~al.}(2019)\citenamefont {Junot},
  \citenamefont {Figueroa-Morales}, \citenamefont {Darnige}, \citenamefont
  {Lindner}, \citenamefont {Soto}, \citenamefont {Auradou},\ and\ \citenamefont
  {Cl{\'e}ment}}]{Junot2019}%
  \BibitemOpen
  \bibfield  {author} {\bibinfo {author} {\bibfnamefont {G.}~\bibnamefont
  {Junot}}, \bibinfo {author} {\bibfnamefont {N.}~\bibnamefont
  {Figueroa-Morales}}, \bibinfo {author} {\bibfnamefont {T.}~\bibnamefont
  {Darnige}}, \bibinfo {author} {\bibfnamefont {A.}~\bibnamefont {Lindner}},
  \bibinfo {author} {\bibfnamefont {R.}~\bibnamefont {Soto}}, \bibinfo {author}
  {\bibfnamefont {H.}~\bibnamefont {Auradou}}, \ and\ \bibinfo {author}
  {\bibfnamefont {E.}~\bibnamefont {Cl{\'e}ment}},\ }\href@noop {} {\bibfield
  {journal} {\bibinfo  {journal} {Europhysics Letters}\ }\textbf {\bibinfo
  {volume} {126}},\ \bibinfo {pages} {44003} (\bibinfo {year}
  {2019})}\BibitemShut {NoStop}%
\bibitem [{\citenamefont {Bondoc-Naumovitz}\ \emph {et~al.}(2023)\citenamefont
  {Bondoc-Naumovitz}, \citenamefont {Laeverenz-Schlogelhofer}, \citenamefont
  {Poon}, \citenamefont {Boggon}, \citenamefont {Bentley}, \citenamefont
  {Cortese},\ and\ \citenamefont {Wan}}]{bondoc2023methods}%
  \BibitemOpen
  \bibfield  {author} {\bibinfo {author} {\bibfnamefont {K.~G.}\ \bibnamefont
  {Bondoc-Naumovitz}}, \bibinfo {author} {\bibfnamefont {H.}~\bibnamefont
  {Laeverenz-Schlogelhofer}}, \bibinfo {author} {\bibfnamefont {R.~N.}\
  \bibnamefont {Poon}}, \bibinfo {author} {\bibfnamefont {A.~K.}\ \bibnamefont
  {Boggon}}, \bibinfo {author} {\bibfnamefont {S.~A.}\ \bibnamefont {Bentley}},
  \bibinfo {author} {\bibfnamefont {D.}~\bibnamefont {Cortese}}, \ and\
  \bibinfo {author} {\bibfnamefont {K.~Y.}\ \bibnamefont {Wan}},\ }\href@noop
  {} {\bibfield  {journal} {\bibinfo  {journal} {Integrative and Comparative
  Biology}\ }\textbf {\bibinfo {volume} {63}},\ \bibinfo {pages} {1485}
  (\bibinfo {year} {2023})}\BibitemShut {NoStop}%
\bibitem [{\citenamefont {Barnkob}\ and\ \citenamefont
  {Rossi}(2020)}]{barnkob2020general}%
  \BibitemOpen
  \bibfield  {author} {\bibinfo {author} {\bibfnamefont {R.}~\bibnamefont
  {Barnkob}}\ and\ \bibinfo {author} {\bibfnamefont {M.}~\bibnamefont
  {Rossi}},\ }\href@noop {} {\bibfield  {journal} {\bibinfo  {journal}
  {Experiments in Fluids}\ }\textbf {\bibinfo {volume} {61}},\ \bibinfo {pages}
  {1} (\bibinfo {year} {2020})}\BibitemShut {NoStop}%
\bibitem [{\citenamefont {Bachimanchi}\ \emph {et~al.}(2022)\citenamefont
  {Bachimanchi}, \citenamefont {Midtvedt}, \citenamefont {Midtvedt},
  \citenamefont {Selander},\ and\ \citenamefont {Volpe}}]{Bachimanchi2022}%
  \BibitemOpen
  \bibfield  {author} {\bibinfo {author} {\bibfnamefont {H.}~\bibnamefont
  {Bachimanchi}}, \bibinfo {author} {\bibfnamefont {B.}~\bibnamefont
  {Midtvedt}}, \bibinfo {author} {\bibfnamefont {D.}~\bibnamefont {Midtvedt}},
  \bibinfo {author} {\bibfnamefont {E.}~\bibnamefont {Selander}}, \ and\
  \bibinfo {author} {\bibfnamefont {G.}~\bibnamefont {Volpe}},\ }\href@noop {}
  {\bibfield  {journal} {\bibinfo  {journal} {Elife}\ }\textbf {\bibinfo
  {volume} {11}},\ \bibinfo {pages} {e79760} (\bibinfo {year}
  {2022})}\BibitemShut {NoStop}%
\bibitem [{\citenamefont {Memmolo}\ \emph {et~al.}(2015)\citenamefont
  {Memmolo}, \citenamefont {Miccio}, \citenamefont {Paturzo}, \citenamefont
  {Caprio}, \citenamefont {Coppola}, \citenamefont {Netti},\ and\ \citenamefont
  {Ferraro}}]{Memmolo2015}%
  \BibitemOpen
  \bibfield  {author} {\bibinfo {author} {\bibfnamefont {P.}~\bibnamefont
  {Memmolo}}, \bibinfo {author} {\bibfnamefont {L.}~\bibnamefont {Miccio}},
  \bibinfo {author} {\bibfnamefont {M.}~\bibnamefont {Paturzo}}, \bibinfo
  {author} {\bibfnamefont {G.~D.}\ \bibnamefont {Caprio}}, \bibinfo {author}
  {\bibfnamefont {G.}~\bibnamefont {Coppola}}, \bibinfo {author} {\bibfnamefont
  {P.~A.}\ \bibnamefont {Netti}}, \ and\ \bibinfo {author} {\bibfnamefont
  {P.}~\bibnamefont {Ferraro}},\ }\href@noop {} {\bibfield  {journal} {\bibinfo
   {journal} {Adv. Opt. Photon.}\ }\textbf {\bibinfo {volume} {7}},\ \bibinfo
  {pages} {713} (\bibinfo {year} {2015})}\BibitemShut {NoStop}%
\bibitem [{\citenamefont {Yajima}\ \emph {et~al.}(2008)\citenamefont {Yajima},
  \citenamefont {Mizutani},\ and\ \citenamefont {Nishizaka}}]{Yajima2008}%
  \BibitemOpen
  \bibfield  {author} {\bibinfo {author} {\bibfnamefont {J.}~\bibnamefont
  {Yajima}}, \bibinfo {author} {\bibfnamefont {K.}~\bibnamefont {Mizutani}}, \
  and\ \bibinfo {author} {\bibfnamefont {T.}~\bibnamefont {Nishizaka}},\
  }\href@noop {} {\bibfield  {journal} {\bibinfo  {journal} {Nature Structural
  \& Molecular Biology}\ }\textbf {\bibinfo {volume} {15}},\ \bibinfo {pages}
  {1119} (\bibinfo {year} {2008})}\BibitemShut {NoStop}%
\bibitem [{\citenamefont {Marumo}\ \emph {et~al.}(2021)\citenamefont {Marumo},
  \citenamefont {Yamagishi},\ and\ \citenamefont {Yajima}}]{Marumo2021}%
  \BibitemOpen
  \bibfield  {author} {\bibinfo {author} {\bibfnamefont {A.}~\bibnamefont
  {Marumo}}, \bibinfo {author} {\bibfnamefont {M.}~\bibnamefont {Yamagishi}}, \
  and\ \bibinfo {author} {\bibfnamefont {J.}~\bibnamefont {Yajima}},\
  }\href@noop {} {\bibfield  {journal} {\bibinfo  {journal} {Communications
  Biology}\ }\textbf {\bibinfo {volume} {4}},\ \bibinfo {pages} {1209}
  (\bibinfo {year} {2021})}\BibitemShut {NoStop}%
\bibitem [{\citenamefont {Rieu}\ \emph {et~al.}(2021)\citenamefont {Rieu},
  \citenamefont {Vieille}, \citenamefont {Radou}, \citenamefont {Jeanneret},
  \citenamefont {Ruiz-Gutierrez}, \citenamefont {Ducos}, \citenamefont
  {Allemand},\ and\ \citenamefont {Croquette}}]{Rieu2021}%
  \BibitemOpen
  \bibfield  {author} {\bibinfo {author} {\bibfnamefont {M.}~\bibnamefont
  {Rieu}}, \bibinfo {author} {\bibfnamefont {T.}~\bibnamefont {Vieille}},
  \bibinfo {author} {\bibfnamefont {G.}~\bibnamefont {Radou}}, \bibinfo
  {author} {\bibfnamefont {R.}~\bibnamefont {Jeanneret}}, \bibinfo {author}
  {\bibfnamefont {N.}~\bibnamefont {Ruiz-Gutierrez}}, \bibinfo {author}
  {\bibfnamefont {B.}~\bibnamefont {Ducos}}, \bibinfo {author} {\bibfnamefont
  {J.-F.}\ \bibnamefont {Allemand}}, \ and\ \bibinfo {author} {\bibfnamefont
  {V.}~\bibnamefont {Croquette}},\ }\href@noop {} {\bibfield  {journal}
  {\bibinfo  {journal} {Science Advances}\ }\textbf {\bibinfo {volume} {7}},\
  \bibinfo {pages} {eabe3902} (\bibinfo {year} {2021})}\BibitemShut {NoStop}%
\bibitem [{\citenamefont {Darnige}\ \emph {et~al.}(2017)\citenamefont
  {Darnige}, \citenamefont {Figueroa-Morales}, \citenamefont {Bohec},
  \citenamefont {Lindner},\ and\ \citenamefont {Cl{\'e}ment}}]{Darnige2017}%
  \BibitemOpen
  \bibfield  {author} {\bibinfo {author} {\bibfnamefont {T.}~\bibnamefont
  {Darnige}}, \bibinfo {author} {\bibfnamefont {N.}~\bibnamefont
  {Figueroa-Morales}}, \bibinfo {author} {\bibfnamefont {P.}~\bibnamefont
  {Bohec}}, \bibinfo {author} {\bibfnamefont {A.}~\bibnamefont {Lindner}}, \
  and\ \bibinfo {author} {\bibfnamefont {E.}~\bibnamefont {Cl{\'e}ment}},\
  }\href@noop {} {\bibfield  {journal} {\bibinfo  {journal} {Review of
  Scientific Instruments}\ }\textbf {\bibinfo {volume} {88}} (\bibinfo {year}
  {2017})}\BibitemShut {NoStop}%
\bibitem [{\citenamefont {Maas}\ \emph {et~al.}(1993)\citenamefont {Maas},
  \citenamefont {Gr{\"u}n},\ and\ \citenamefont {Papantoniou}}]{maas1993}%
  \BibitemOpen
  \bibfield  {author} {\bibinfo {author} {\bibfnamefont {H.}~\bibnamefont
  {Maas}}, \bibinfo {author} {\bibfnamefont {A.}~\bibnamefont {Gr{\"u}n}}, \
  and\ \bibinfo {author} {\bibfnamefont {D.}~\bibnamefont {Papantoniou}},\
  }\href@noop {} {\bibfield  {journal} {\bibinfo  {journal} {Experiments in
  Fluids}\ ,\ \bibinfo {pages} {133}} (\bibinfo {year} {1993})}\BibitemShut
  {NoStop}%
\bibitem [{\citenamefont {Prasad}(2000)}]{Prasad2000}%
  \BibitemOpen
  \bibfield  {author} {\bibinfo {author} {\bibfnamefont {A.~K.}\ \bibnamefont
  {Prasad}},\ }\href@noop {} {\bibfield  {journal} {\bibinfo  {journal}
  {Experiments in Fluids}\ }\textbf {\bibinfo {volume} {29}},\ \bibinfo {pages}
  {103} (\bibinfo {year} {2000})}\BibitemShut {NoStop}%
\bibitem [{\citenamefont {Kim}\ \emph {et~al.}(2011)\citenamefont {Kim},
  \citenamefont {Gro{\ss}e}, \citenamefont {Elsinga},\ and\ \citenamefont
  {Westerweel}}]{kim2011}%
  \BibitemOpen
  \bibfield  {author} {\bibinfo {author} {\bibfnamefont {H.}~\bibnamefont
  {Kim}}, \bibinfo {author} {\bibfnamefont {S.}~\bibnamefont {Gro{\ss}e}},
  \bibinfo {author} {\bibfnamefont {G.~E.}\ \bibnamefont {Elsinga}}, \ and\
  \bibinfo {author} {\bibfnamefont {J.}~\bibnamefont {Westerweel}},\
  }\href@noop {} {\bibfield  {journal} {\bibinfo  {journal} {Experiments in
  Fluids}\ }\textbf {\bibinfo {volume} {51}},\ \bibinfo {pages} {395} (\bibinfo
  {year} {2011})}\BibitemShut {NoStop}%
\bibitem [{\citenamefont {Drescher}\ \emph {et~al.}(2009)\citenamefont
  {Drescher}, \citenamefont {Leptos},\ and\ \citenamefont
  {Goldstein}}]{Drescher2009}%
  \BibitemOpen
  \bibfield  {author} {\bibinfo {author} {\bibfnamefont {K.}~\bibnamefont
  {Drescher}}, \bibinfo {author} {\bibfnamefont {K.~C.}\ \bibnamefont
  {Leptos}}, \ and\ \bibinfo {author} {\bibfnamefont {R.~E.}\ \bibnamefont
  {Goldstein}},\ }\href@noop {} {\bibfield  {journal} {\bibinfo  {journal}
  {Review of scientific instruments}\ }\textbf {\bibinfo {volume} {80}}
  (\bibinfo {year} {2009})}\BibitemShut {NoStop}%
\bibitem [{\citenamefont {Cheung}\ \emph {et~al.}(2005)\citenamefont {Cheung},
  \citenamefont {Ng},\ and\ \citenamefont {Zhang}}]{Cheung2005}%
  \BibitemOpen
  \bibfield  {author} {\bibinfo {author} {\bibfnamefont {K.}~\bibnamefont
  {Cheung}}, \bibinfo {author} {\bibfnamefont {W.}~\bibnamefont {Ng}}, \ and\
  \bibinfo {author} {\bibfnamefont {Y.}~\bibnamefont {Zhang}},\ }\href@noop {}
  {\bibfield  {journal} {\bibinfo  {journal} {Flow Measurement and
  Instrumentation}\ }\textbf {\bibinfo {volume} {16}},\ \bibinfo {pages} {295}
  (\bibinfo {year} {2005})}\BibitemShut {NoStop}%
\bibitem [{\citenamefont {Peterson}\ \emph {et~al.}(2012)\citenamefont
  {Peterson}, \citenamefont {Regaard}, \citenamefont {Heinemann},\ and\
  \citenamefont {Sick}}]{Peterson2012}%
  \BibitemOpen
  \bibfield  {author} {\bibinfo {author} {\bibfnamefont {K.}~\bibnamefont
  {Peterson}}, \bibinfo {author} {\bibfnamefont {B.}~\bibnamefont {Regaard}},
  \bibinfo {author} {\bibfnamefont {S.}~\bibnamefont {Heinemann}}, \ and\
  \bibinfo {author} {\bibfnamefont {V.}~\bibnamefont {Sick}},\ }\href@noop {}
  {\bibfield  {journal} {\bibinfo  {journal} {Opt. Express}\ }\textbf {\bibinfo
  {volume} {20}},\ \bibinfo {pages} {9031} (\bibinfo {year}
  {2012})}\BibitemShut {NoStop}%
\bibitem [{\citenamefont {Lertvilai}\ \emph {et~al.}(2021)\citenamefont
  {Lertvilai}, \citenamefont {Roberts},\ and\ \citenamefont
  {Jaffe}}]{Lertvilai2021}%
  \BibitemOpen
  \bibfield  {author} {\bibinfo {author} {\bibfnamefont {P.}~\bibnamefont
  {Lertvilai}}, \bibinfo {author} {\bibfnamefont {P.~L.}\ \bibnamefont
  {Roberts}}, \ and\ \bibinfo {author} {\bibfnamefont {J.~S.}\ \bibnamefont
  {Jaffe}},\ }\href@noop {} {\bibfield  {journal} {\bibinfo  {journal} {Journal
  of Atmospheric and Oceanic Technology}\ }\textbf {\bibinfo {volume} {38}},\
  \bibinfo {pages} {1143 } (\bibinfo {year} {2021})}\BibitemShut {NoStop}%
\bibitem [{\citenamefont {Finney}\ \emph {et~al.}(2022)\citenamefont {Finney},
  \citenamefont {Skrodzki}, \citenamefont {Peskosky}, \citenamefont {Burger},
  \citenamefont {Nees}, \citenamefont {Krushelnick},\ and\ \citenamefont
  {Jovanovic}}]{finney2022}%
  \BibitemOpen
  \bibfield  {author} {\bibinfo {author} {\bibfnamefont {L.~A.}\ \bibnamefont
  {Finney}}, \bibinfo {author} {\bibfnamefont {P.~J.}\ \bibnamefont
  {Skrodzki}}, \bibinfo {author} {\bibfnamefont {N.}~\bibnamefont {Peskosky}},
  \bibinfo {author} {\bibfnamefont {M.}~\bibnamefont {Burger}}, \bibinfo
  {author} {\bibfnamefont {J.}~\bibnamefont {Nees}}, \bibinfo {author}
  {\bibfnamefont {K.}~\bibnamefont {Krushelnick}}, \ and\ \bibinfo {author}
  {\bibfnamefont {I.}~\bibnamefont {Jovanovic}},\ }\href@noop {} {\bibfield
  {journal} {\bibinfo  {journal} {Scientific Reports}\ }\textbf {\bibinfo
  {volume} {12}},\ \bibinfo {pages} {17205} (\bibinfo {year}
  {2022})}\BibitemShut {NoStop}%
\bibitem [{\citenamefont {Foster}\ and\ \citenamefont
  {Smyth}(1980)}]{foster1980}%
  \BibitemOpen
  \bibfield  {author} {\bibinfo {author} {\bibfnamefont {K.}~\bibnamefont
  {Foster}}\ and\ \bibinfo {author} {\bibfnamefont {R.}~\bibnamefont {Smyth}},\
  }\href {\doibase 10.1128/MR.44.4.572-630.1980} {\bibfield  {journal}
  {\bibinfo  {journal} {Microbiological Reviews}\ }\textbf {\bibinfo {volume}
  {44}},\ \bibinfo {pages} {572} (\bibinfo {year} {1980})}\BibitemShut
  {NoStop}%
\bibitem [{\citenamefont {Allan}\ \emph {et~al.}(2024)\citenamefont {Allan},
  \citenamefont {Caswell}, \citenamefont {Keim}, \citenamefont {van~der Wel},\
  and\ \citenamefont {Verweij}}]{trackpy}%
  \BibitemOpen
  \bibfield  {author} {\bibinfo {author} {\bibfnamefont {D.~B.}\ \bibnamefont
  {Allan}}, \bibinfo {author} {\bibfnamefont {T.}~\bibnamefont {Caswell}},
  \bibinfo {author} {\bibfnamefont {N.~C.}\ \bibnamefont {Keim}}, \bibinfo
  {author} {\bibfnamefont {C.~M.}\ \bibnamefont {van~der Wel}}, \ and\ \bibinfo
  {author} {\bibfnamefont {R.~W.}\ \bibnamefont {Verweij}},\ }\href {\doibase
  10.5281/zenodo.11522100} {\enquote {\bibinfo {title} {soft-matter/trackpy:
  v0.6.3},}\ } (\bibinfo {year} {2024})\BibitemShut {NoStop}%
\bibitem [{\citenamefont {Crocker}\ and\ \citenamefont
  {Grier}(1996)}]{crocker1996methods}%
  \BibitemOpen
  \bibfield  {author} {\bibinfo {author} {\bibfnamefont {J.~C.}\ \bibnamefont
  {Crocker}}\ and\ \bibinfo {author} {\bibfnamefont {D.~G.}\ \bibnamefont
  {Grier}},\ }\href@noop {} {\bibfield  {journal} {\bibinfo  {journal} {Journal
  of colloid and interface science}\ }\textbf {\bibinfo {volume} {179}},\
  \bibinfo {pages} {298} (\bibinfo {year} {1996})}\BibitemShut {NoStop}%
\bibitem [{\citenamefont {Delplace}(2018)}]{delplace2018laminar}%
  \BibitemOpen
  \bibfield  {author} {\bibinfo {author} {\bibfnamefont {F.}~\bibnamefont
  {Delplace}},\ }\href@noop {} {\bibfield  {journal} {\bibinfo  {journal} {Open
  Access J. Math. Theor. Phys.}\ }\textbf {\bibinfo {volume} {1}},\ \bibinfo
  {pages} {198} (\bibinfo {year} {2018})}\BibitemShut {NoStop}%
\end{thebibliography}%

\end{document}